\documentclass[aps,prl,twocolumn,groupedaddress,nofootinbib,showpacs]{revtex4}
\usepackage[dvips]{graphicx}
\newif\ifdraft
\drafttrue
\newif\ifpreprint
\preprinttrue

\def\spa#1.#2{\left\langle#1\,#2\right\rangle}
\def\spb#1.#2{\left[#1\,#2\right]}

\def\e{\epsilon}

\newcommand{\eq}{\begin{equation}}
\newcommand{\eqe}{\end{equation}}
\newcommand{\eqa}{\begin{eqnarray}}
\newcommand{\eqae}{\end{eqnarray}}

\newbox\charbox
\newbox\slabox
\def\s#1{{      
        \setbox\charbox=\hbox{$#1$}
        \setbox\slabox=\hbox{$/$}
        \dimen\charbox=\ht\slabox
        \advance\dimen\charbox by -\dp\slabox
        \advance\dimen\charbox by -\ht\charbox
        \advance\dimen\charbox by \dp\charbox
        \divide\dimen\charbox by 2
        \raise-\dimen\charbox\hbox to \wd\charbox{\hss/\hss}
        \llap{$#1$}
}}

\begin{document}

\title{
\ifpreprint
\hbox{\rm \small 
$\null$ \hskip 2 cm \hfill PUPT-2450 \hskip 9.5 cm \hfill  MCTP-13-20} 
\fi
Consistency conditions from generalized-unitarity }
 
\author{Yu-tin Huang$^{a}$, David A. McGady$^b$
}

\affiliation{
${}^a$Michigan Center for Theoretical Physics, Department of Physics, University of Michigan, Ann Arbor, MI 48109, USA\\
${}^b$ Department of Physics, Princeton University, Princeton, NJ 08544}

\begin{abstract} 
In the modern on-shell approach, the perturbative S-matrix is constructed iteratively using on-shell building blocks with manifest unitarity. As only gauge invariant quantities enter in the intermediate steps, the notion of gauge anomaly is absent. In this letter, we rephrase the anomaly cancellation conditions in a purely on-shell language. We demonstrate that while the unitarity-methods automatically lead to a unitary S-matrix, the rational terms that are required to enforce locality, invariably give rise to inconsistent factorization channels in chiral theories. In four-dimensions, the absence of such inconsistencies implies the vanishing of the cubic Casimir of the gauge group. In six-dimensions, if the symmetric trace of four generators does not vanish, the rational term develops a factorization channel revealing a new particle in the spectrum: the two-form of the Green-Schwarz mechanism. Thus in the purely on-shell construction, the notion of gauge-anomaly is replaced by the difficulty to consistently impose locality on the unitary S-matrix.
\end{abstract}

\pacs{04.65.+e, 11.15.Bt, 11.30.Pb, 11.55.Bq \hspace{1cm}}

\maketitle

An intriguing difference between the traditional Lagrangian definition of perturbative quantum field theory (QFT) and the modern analytic S-matrix program, is the role of gauge symmetry. Where as gauge invariance is crucial in determining the Lagrangian and ensures unitarity of the perturbative S-matrix, such notions are completely absent in the modern on-shell approach. In the latter approach, given the free-spectrum of the theory, the lowest-multiplicity non-trivial S-matrix can be determined completely from the global symmetries of the theory. Using factorization~\cite{bcfwref} as well as unitarity constraints~\cite{UnitarityMethod}, the entire perturbative S-matrix can then be iteratively constructed from that of the lowest order. Such an approach has led to tremendous progress in the computation of high loop-order corrections in four-dimensional super Yang-Mills~\cite{5LSYM}, supergravity~\cite{4DSG}, higher-dimensional super Yang-Mills~\cite{5DSYM}, as well as the determination of all-loop planar integrand of maximal super Yang-Mills~\cite{SingleCut3}. Since the building blocks that enter the iterative process are completely on-shell, gauge invariance becomes a notion that is devoid of substance.      

The fact that the physical observables of a QFT can be constructed without the utterance of gauge symmetry, leads us to ask how consistency constraints traditionally \emph{imposed} by the requirement of gauge anomaly cancellation, arises in such on-shell constructions. Establishment of such constraint without knowledge of the interaction Lagrangian
, becomes crucial in light of the large class of supersymmetric Chern-Simons matter theories~\cite{BLG, ABJM} whose Lagrangian has been constructed only in the past five years, although their S-matrix elements can be determined independently~\cite{Bargheer:2010hn, Huang:2010rn}. 

In this letter, we address the following question: starting with a theory of chiral fermions, as we construct loop-amplitudes through the on-shell program, how do we see that the theory is sick? Tree-level amplitudes of chiral fermions are perfectly well defined. Through general unitarity methods, one necessarily obtains a unitary S-matrix. Superficially, chiral gauge theories should have perfectly sensible loop amplitudes. However, while the S-matrix is manifestly unitary, it contains spurious non-local poles. To ensure that the final result is both unitary and local, one is forced to introduce non cut-constructible rational terms to cancel the spurious poles. We will demonstrate that for chiral fermion loops, cancellation of these spurious singularities induces new factorization channels. In four-dimensions, such factorization channels are inconsistent and thus must cancel. The constraint imposed by such cancellation is precisely the vanishing of the cubic Casimir of the gauge group. In six-dimensions, if the symmetric trace of the four generators does not vanish, the new induced factorization channel reveals the presence of a new particle in the theory: the two-form in the Green-Schwarz (GS) mechanism~\cite{GS}.
\section{A prelude in four-dimensions}
Unitarity methods naturally cast one-loop amplitudes into a basis of scalar integrals whose coefficients depend on the theory at hand. Here, we consider the fermion-loop contribution to the single trace one-loop four-gluon amplitude. For later convenience we give the scalar-integral coefficients originating from two distinct fermion helicities separately:
\eqa
\nonumber -\frac{t^4s^2}{u^4}\;\;\vcenter{\hbox{\includegraphics[scale=0.45]{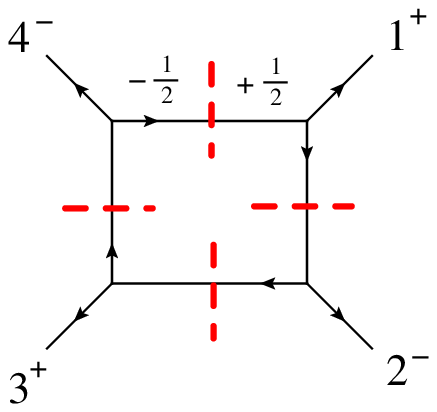}}} &&-\frac{s^4t^2}{u^4}\quad\vcenter{\hbox{\includegraphics[scale=0.45]{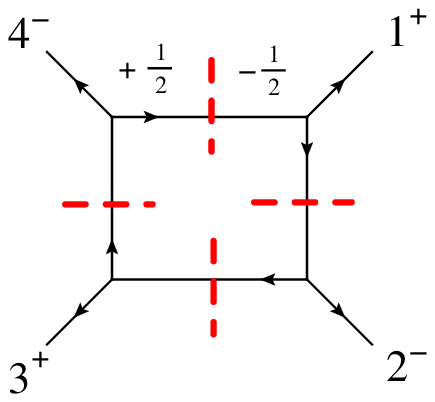}}} \\
\nonumber  \frac{t^4s}{u^4} \;\quad\vcenter{\hbox{\includegraphics[scale=0.5]{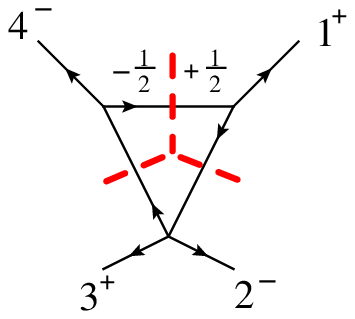}}}\;&&\;\;\; \frac{t^2s^3}{u^4}\quad\;\,\vcenter{\hbox{\includegraphics[scale=0.5]{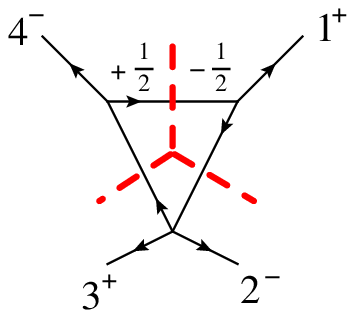}}}\\
\nonumber \vcenter{\hbox{\includegraphics[scale=0.45]{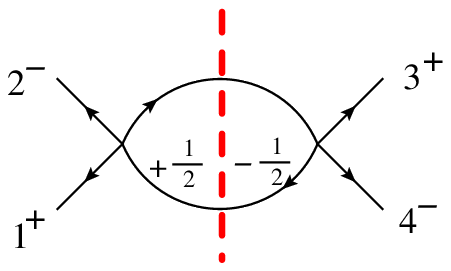}}}&&\quad\quad\quad\quad \vcenter{\hbox{\includegraphics[scale=0.45]{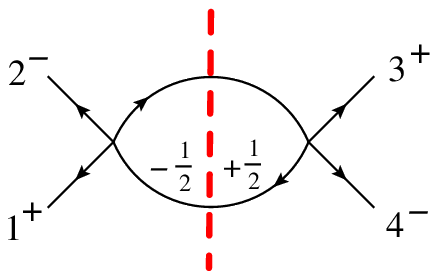}}}\\
\nonumber \;\frac{t(su-6st-2ut)}{6u^3}&& \quad\quad  \frac{t(4 s^2 + 2 t^2 - 7 s u)}{6u^3}\,.\\
\eqae
In the above, we've indicated the helicities of the fermions crossing the unitarity cut, denoted by the (red) dashed lines, and $s=(k_1+k_2)^2$, $t=(k_2+k_3)^2$, $u=(k_1+k_3)^2$. Note that the triangle- and box-integral coefficients are such that the IR-divergence cancels, which is necessary due to the absence of tree-level processes for a fermion in background gauge field. 

The the parity-even part of the fermion-loop amplitude simply corresponds to the sum of the two distinct helicity configurations:
\eqa
&&\nonumber  \frac{A^{\rm even}(1^+2^-3^+4^-)}{A^{\rm tree}}=-\frac{st(s^2+t^2)}{2u^4}\bigg(\log\left(\frac{t}{s}\right)^2+\pi^2\bigg)\\
&& +\left[\left(\frac{s-t}{3u}-\frac{st(s-t)}{u^3}\right)\right]\log\left(\frac{s}{t}\right)-\frac{(-s)^{-\e}+(-t)^{-\e}}{3\epsilon}\nonumber\\
&&+R^{\rm even}_{(1,2,3,4)}\,,
\label{even}
\eqae
where we've included a term $R^{\rm \,even}$ representing possible rational terms that are undetectable from unitarity cuts, and $A^{\rm tree}$ is the tree-level amplitude. The rational term can be determined from imposing locality. To see this, note that poles in the $u$-channel are ubiquitous throughout Eq.(\ref{even}), which cannot have a local interpretation due to the color-ordering. As $u\rightarrow0$, Eq.(\ref{even}) behaves as:
\eq
\left(Eq.(\ref{even})-R^{\rm even}_{(1,2,3,4)}\right)\bigg|_{u\rightarrow0}=-\frac{s^2}{u^2}-\frac{s}{u}+\mathcal{O}(u^0)\,.
\eqe
Locality requires $R^{\rm even}_{(1,2,3,4)}$ to cancel these spurious poles . Dimension-counting and cyclic invariance uniquely fixes it to be,
\eq
R^{\rm even}_{(1,2,3,4)}=-\frac{st}{u^2} \,.
\label{nonCR}
\eqe
Substituting Eq.(\ref{nonCR}) into Eq.(\ref{even}) reproduces known results in QCD~\cite{BernQCD}. Note that since the amplitude has an $A^{\rm tree}$ pre factor, the presence of a rational term can potentially introduce new residues on the physical poles of the tree-amplitude. However, due to the $st$ factor in the numerator of Eq.(\ref{nonCR}), the residue vanishes. 

We now turn to the parity-odd part of the fermion-loop, which is only present for chiral fermions. It is simply given by the difference of the two helicity configurations: 
\eqa
\nonumber  &&\frac{A^{\rm odd}(1^+2^-3^+4^-)}{A^{\rm tree}}=-\frac{st(s-t)}{2u^3}\bigg(\log\left(\frac{t}{s}\right)^2+\pi^2\bigg)\\
&&\quad\quad\quad-\left(\frac{2st}{u^2}\right)\log\left(\frac{-s}{-t}\right)+R^{\rm odd}_{(1,2,3,4)} \,.
\label{odd}
\eqae
First, note that the amplitude is cyclic invariant up to a sign, which is due to the use of helicity basis. As with the parity-even combination, there are spurious $u$-channel poles in Eq.(\ref{odd}):
\eq
\left(Eq.(\ref{odd})-R^{\rm odd}_{(1,2,3,4)}\right)|_{u\rightarrow0}=-\frac{s}{u}+\mathcal{O}(u^0)\,.
\eqe 
Locality again requires such spurious poles to be canceled by $R^{\rm odd}_{(1,2,3,4)}$. Taking into account the fact that the amplitude attains a minus sign under cyclic shift, the requisite parity-odd rational term is:
\eq
A^{\rm tree}R^{\rm odd}_{(1,2,3,4)}=A^{\rm tree}\frac{s-t}{2u} = \langle24\rangle^2 [13]^2\frac{s-t}{2stu} \,.
\label{CR}
\eqe
However, as is plain from Eq.(\ref{CR}), this new parity-odd rational term has non-trivial contributions to the $s$- and $t$-channel residues. This contrasts sharply with the $R^{\rm \,even}_{(1,2,3,4)}$. Herein lay the seeds of inconsistencies in parity-violating gauge theories: the rational terms that are required for locality in the parity-odd amplitude, introduce \emph{new} corrections to residues on the $s$- and $t$- poles. This is inconsistent. To see why, note that as the residue of the pole has mass-dimension two, it can only factorize into three-point functions, each with mass-dimension one. However, there are exactly two \emph{unique} mass-dimension one three-point amplitudes involving two gauge fields. They are the Yang-Mills three-point amplitudes, and are \emph{entirely fixed} by Poincare invariance. They do not have one-loop corrections. Thus the absence of acceptable residues implies that such factorization channel is inconsistent.

Requiring these inconsistent factorization channels to be absent from the amplitude constrains the theory. To see how, note that there are 6 single-trace color structures at four-points and one-loop. Four of these contain such excess residue in the physical $s$-channel. Their rational terms sum to,
\eqa
A^{\rm tree}\frac{s-t}{2u}\left(tr[T^1T^4T^3T^2]-tr[T^1T^2T^3T^4]+(1\leftrightarrow2)\right)
\eqae
Thus we see that the problematic residues from the rational terms exactly cancel if the group-theory factor vanishes:
\eqa
&&tr[T^1T^2T^3T^4]-tr[T^1T^4T^3T^2]+(1\leftrightarrow2)\nonumber\\
\nonumber&=&d^{1a4}f^{23}\,_a+d^{13a}f^{24}\,_a+d^{1a2}f^{34}\,_a+(1\leftrightarrow2)=0\,.\\
\eqae
Since the symmetry property of each term is distinct, the constraint is satisfied only if each term is individually zero. Thus imposing unitarity and locality, one arrives at the following constraint on the group-theory factor: 
\eq
d^{abc}f^{de}\,_a=0\,.
\label{4DConstraint}
\eqe
This is nothing but the anomaly cancellation condition of the non-abelian box anomaly! In summary, in using unitarity methods to construct one-loop scattering amplitudes, one encounters obstacles in implementing locality if there are chiral fermion loops. Such obstruction ceases to exist if the theory has vanishing $d_{abc}$.

\section{The 6D rational term and the GS two-form}
We now consider the one-loop four-point amplitude in 6D chiral gauge-theory. The little-group in six-dimensions is SO(4)=SU(2)$\times$SU(2), and for the parity-odd contribution, we again take the difference between $(\frac{1}{2},0)$ and $(0,\frac{1}{2})$ fermion in the loop. To obtain the scalar integral coefficients, we utilize six-dimensional spinor-helicity formalism~\cite{Donal} as well as generalized unitarity-methods~\cite{Huang6D}.  Explicit computation gives the following coefficients for the scalar box, triangle and bubble integrals respectively:
\eqa
\nonumber C_4&=&\frac{(s-t)}{6u^2}F^{(4)},\;\; C_{3s}=-\frac{(s-t)}{6tu^2}F^{(4)},\\
\nonumber C_{3t}&=&-\frac{(s-t)}{6su^2}F^{(4)},\;\; C_{2s}=\frac{F^{(4)}}{stu},\;\;C_{2t}=-\frac{F^{(4)}}{stu} \, ,
\eqae
The function $F^{(4)}$ is explicitly given as:\footnote{The on-shell form of the wedge product of three field strengths is simply
\eqa
&&F_1\wedge F_2\wedge F_3 
= \left( \langle1_a|2_{\dot{b}}]\langle2_b|3_{\dot{c}}]\langle3_c|1_{\dot{a}}]+\langle2_b|1_{\dot{a}}]\langle1 _a|3_{\dot{c}}]\langle3_c|2_{\dot{b}}] \right)\,. \nonumber
\eqae} 
\eqa
\nonumber F^{(4)}\equiv\langle 4_{d}|p_2p_3|4_{\dot{d}}]F^3_{(123)}+ (\sigma_i){\rm cyclic}\,,
\eqae
where $+$cyclic indicates the sum over remaining three cyclic permutations,  $\sigma_i$ is the signature of each permutation, and $F^3_{(ijk)}\equiv F_i\wedge F_j\wedge F_k$. Explicitly evaluating the scalar integrals yield the parity-odd portion of the chiral fermion contribution to the four-gluon amplitude:
\eqa
\nonumber \frac{A^{\rm \,odd}(1,2,3,4)}{F^{(4)}}&=&\frac{(t-s)\left(\pi^2+\log[s/t]^2\right)}{12u^3}+\frac{\log[t/s]}{3u^2}\\
&&+\frac{s-t}{18stu} + R^{\rm odd}_{(1,2,3,4)}\,, \label{odd1}
\eqae
where again $R^{\rm odd}_{(1,2,3,4)}$ represents the possible cut-free rational term. Firstly, note that the ultraviolet (UV) divergences explicitly canceled, just as the infrared divergence cancelled in four-dimensions. The absence of UV-divergences in the parity-odd amplitude must hold, as there are no local operators available as viable counter-terms. Secondly, rational terms are already present in the cut-constructible anwser. This is a subtle difference from the previous $D=4$ analysis, and only appears in higher-dimensions. The origin of this is due to the fact that while only scalar bubbles are UV-divergent in $D=4$, for higher dimensions all $n\leq D/2$-gon scalar integrals are UV-divergent. The cancellation of UV-divergence then invariably leaves behind a rational term. For example in $D=6$ the bubble- and triangle-integrals, in dimensional regularization, are given as,
\eqa
I_3[K^2]&=&\frac{1}{2 \e} + \frac{1}{2} \left(3 - \gamma_E - \log[K^2]\right),\; \nonumber\\
I_2[K^2]&=&-\frac{K^2}{6 \e} + \frac{K^2}{18} \left(-8 +3\gamma_E +3\log[K^2]\right)\,, \label{Integrals}
\eqae
where $K^2$ is the unique kinematic invariant of the integral. As one can see, the cancellation of UV-divergences inevitably lead to a nontrivial rational term. Note that while the scalar integrals contain divergences and require regularization, the amplitude is finite and any result derived from the analysis of the amplitude will be scheme independent.  

Again the ubiquitous presence of $u$-channel poles requires us to ensure that the residue of this pole, which is \emph{spurious} for this ordering, must vanish to ensure locality. As $u\rightarrow0$ one finds: 
\eq
\left(Eq.(\ref{odd1})-R^{\rm odd}_{(1,2,3,4)}\right)\bigg|_{u\rightarrow0}=-\frac{1}{18tu}+\mathcal{O}(u^0)\,. \label{6DUres}
\eqe
Note that although the $u$-channel is spurious in two of the six orderings, it is also present in the remaining four, 
due to the presence of rational terms arising from the cut-constructible part. This is the non-trivial consequence that was previously alluded to. It is straightforward to check that the leading $u\rightarrow0$ behavior of all orderings are identical to Eq.(\ref{6DUres}), and thus for the full color-dressed amplitude, with $R=0$, the leading $u\rightarrow0$ behavior is given as: 
\eq\label{6DFac}
\mathcal{A}|_{u\rightarrow0}=-\frac{1}{18 ut}sTr(1234)+\mathcal{O}(u^0)
\eqe
where $sTr(1234)$ is the symmetric trace of the four generators. Thus for the absence of factorization poles one must have:
\eq
sTr(1234)=0\,.
\label{6Danom}
\eqe
This reproduces the standard anomaly cancellation condition in six-dimensions. If Eq.(\ref{6Danom}) is not satisfied, then one must give a physical interpretation for this new factorization pole. Here, un-like D=4, the mass dimension of this residue is 4, implying that it can factorize into three-point functions that have mass-dimension 2.  Again possible three-point amplitudes are highly constrained by Lorentz invariance, and it can be shown that the only possible dimension 2 amplitudes involving two vector fields is the three-point coupling of a graviton, a scalar, or a two-form to the vectors. Only the latter allows for parity odd-coupling~\cite{YWD}. In other words, in the event that Eq.(\ref{6Danom}) is not satisfied, the factorization pole implies the presence of a new particle in the spectrum: the two-form for the GS mechanism~\cite{GS}.

However this is not the end of the story, since in order for eq.(\ref{6DFac}) to truly correspond to the singularity associated with the exchange of a two-form, the symmetric trace must factorize. This is one of the well known conditions for GS mechanism to apply. However, when the $sTr(1234)$ factorizes, it factorizes into three distinct double trace structure:
\eq\label{TrGen}
sTr(1234)\rightarrow tr(t_2t_4)tr(t_1t_3)+{\rm cyclic(123)}
\eqe
where $t_i$ are the generators in the fundamental representation. Note that only the first term in eq.(\ref{TrGen}) is consistent with an $u$-channel exchange, and the latter still represents inconsistent residues. Thus locality again demands us to add additional rational terms to cancel the inconsistent residues. Again, symmetry properties of the trace structure uniquely fixes the color dressed rational term to be: 
\eq
\mathcal{R}^{\rm odd}=F^{(4)}\left[tr(t_2t_4)tr(t_1t_3)\frac{t-s}{18stu}+{\rm cyclic(123)}\right]\,.
\label{6DR}
\eqe

Remarkably, $\mathcal{R}^{\rm odd}$ is precisely the combination of the anomalous rational term of the Feynman-diagram calculation  and the tree-diagram from GS mechanism. Using integral reduction on the Feynman-diagram loop-integral, one obtains the following parity-odd rational term:\footnote{We use normalization such that the overall factor $1/(4\pi)^3$ is 1.}
\eq
R_{\rm 6D}^{anom}= R^{anom}_{234} + {\rm cyclic}(1234)\,, 
\label{6Da}
\eqe
where:
\eqa
\nonumber R^{anom}_{234} &=& -\frac{1}{18}\left(\frac{(\e_1\cdot k_2)}{s}+\frac{(\e_1\cdot k_3)}{u}+\frac{(\e_1\cdot k_4)}{t}\right)F^3_{(234)}\,.
\eqae
We did not present the expression for distinct orderings since Eq.(\ref{6Da}) is manifestly permutation invariant, as expected. One can easily check that under a gauge transformation $\epsilon_i\rightarrow \epsilon_i+k_i$, Eq.(\ref{6Da}) is anomalous. Working in the fundamental representation and combining with the contribution from the GS mechanism for $tr(t_1t_3)tr(t_2t_4)$, one finds the following gauge invariant combination for this color-factor : 
\eqa\label{Feyn}
\nonumber &&\frac{-1}{18stu}\left[F^3_{(234)}(tu(\e_1\cdot k_2)+su(\e_1\cdot k_4)-2st(\e_1\cdot k_3))\right.\\
\nonumber &&\quad\quad +F^3_{(341)}(tu(\e_2\cdot k_1)+su(\e_2\cdot k_3)-2st(\e_2\cdot k_4))\\
\nonumber &&\quad\quad +F^3_{(412)}(tu(\e_3\cdot k_4)+su(\e_3\cdot k_2)-2st(\e_3\cdot k_1))\\
\nonumber&&\quad\quad\left. +F^3_{(123)}(tu(\e_4\cdot k_3)+su(\e_4\cdot k_1)-2st(\e_4\cdot k_2))\right]\,.\\
\eqae
Converting Eq.(\ref{Feyn}) to on-shell form one finds exactly that of Eq.(\ref{6DR})!

In conclusion, in applying unitarity methods to construct one-loop amplitudes, enforcing locality on chiral fermion loops gives rise to new factorization channels. In $D=4$, such factorization channels lead to inconsistent residues, whose cancellation reproduces the anomaly cancellation conditions. In $D=6$, the absence of spurious singularities requires either constraints which are precisely the well known anomaly cancellation conditions, or the introduction of rational terms to cancel the spurious singularities, leaving behind the physical factorization channels. The new channels then reflect the presence of a new particle in the spectrum: the GS two-form. Furthermore, this unique rational term is precisely the gauge invariant rational term that arrises form the combination of the parity-odd one-loop anomalous rational term and the contribution from the GS-mechanism, computed from Feynman rules. Thus starting with a chiral-gauge theory, imposing locality on the one-loop amplitude directly gives us the complete GS-contribution. In a sense there are no ``gauge anomalies" per se. There are only rational terms in amplitudes, needed to enforce locality in amplitudes built from unitarity methods. 
When present, these rational terms make it impossible to have massless vectors for in $D=4$, whereas they force the existence of new degrees of freedom in higher-dimensions.\footnote{Note that rational terms also play a role in global anomalies, which has been recently discussed in the context of supergravity amplitudes~\cite{Radu}.} From this point of view, as rational terms only appear in even-dimensions at one-loop, and can appear in parity odd-amplitudes beginning at $n=D/2-1$-points, these are the places where such inconsistencies can arise in general, in agreement with the usual gauge-anomaly analysis. Finally, just as the lowest-multiplicity S-matrix can be uniquely determined, so can the parity-odd rational term as we have demonstrated. It will be interesting to see what kind of recursion one can set up to obtain all higher-multiplicity counterparts.
\vskip .3 cm 

We are extremely grateful to Nima Arkani-Hamed for the suggestion and discussion of this problem, Lance Dixon for the useful discussions on the role of rational terms, H. Johansson and Z. Bern for many fruitful discussions. This research was supported by the US DoE grant DE-SC0007859, and DE--FG03--91ER40662.

\end{document}